# Low-cost photoplethysmograph solutions using the Raspberry Pi

Tamás Nagy*, Zoltan Gingl*
* Department of Technical Informatics, University of Szeged, Hungary
nag.tams@gmail.com, gingl@inf.u-szeged.hu

*Abstract*—Photoplethysmography is a prevalent, non-invasive heart monitoring method. In this paper an implementation of photoplethysmography on the Raspberry Pi is presented. Two modulation techniques are discussed, which make possible to measure these signals by the Raspberry Pi, using an external sound card as A/D converter. Furthermore, it is shown, how can digital signal processing improve signal quality. The presented methods can be used in low-cost cardiac function monitoring, in telemedicine applications and in education as well, since cheap and current hardware are used. Full documentation and open-source software for the measurement available: http://www.noise.inf.u-szeged.hu/Instruments/raspberryplet/

## I. Introduction

Photoplethysmography is a current, simple, non-invasive method for monitoring cardiac function [1-5]. Although it can't be used directly to get the absolute blood pressure, the heart beats can be detected regularly. It is possible to get much more information from the time between heart beats, called RR intervals, by signal processing. These hold information about the autonomic nervous system and respiration [5-6].

Our aim was to implement a widely available method for plethysmographic measurements, which is not just usable in low-cost cardiac function monitoring, but also could be built easily and performed at home. Towards that, the measurement device has to be small, portable and reproducible. Hence, the usage of cheap and current means is a very important aspect.

Plethysmographic measurement methods, which have the properties suggested above, are already exist, e. g. for smart phones [6]. Accuracy and reliability are also very important aspects, since it is a medical application. For this reason, it is helpful, to implement the measurement method on a traceable and well known device. In this point of view, getting much more information by signal processing is also very useful. As mentioned above, heart beats can be detected using plethysmograph signals, and many indicators could be calculated from the time intervals between these heart beats. The presented methods should complete a clinical validation, which would also improve reliability.

The device, which is used for this measurement, has to own all of the suggested properties. Portable computers are preferred, like notebooks, netbooks and tablets. Besides those, smart phones, and single-board computers can also be used for this task. In this paper, we show the implementation of a plethysmographic measurement method on the Raspberry Pi, this very common and low-price single-board computer [7]. We note that this measurement method is portable to the above-mentioned, other devices easily.

Since the Raspberry Pi has no analogue input, an external A/D converter is needed. One solution is, to use the PiFace, which has analogue inputs, and communicates with the Raspberry Pi, using the GPIO [8]. The use of this arrangement would lead to a complicated external circuit and measurement software. Moreover, the cost of the PiFace is about as much, as a Raspberry Pi. Another solution is, to use the sound card as A/D converter, which technique insures much simpler, smaller hardware and software, and also costs less [9]. What is even more important, it is portable to almost any platform and operating system. In this article, the second method will be presented, which uses the sound card.

## II. Solutions

Plethysmographic measurement methods are based on the changes of blood volume in an organ. During the cardiac circle the blood pressure is varying, which causes these blood volume changes. The hemoglobin absorbs infrared light, so these changes can be measured using an infrared LED and a phototransistor [1]. In this paper we used reflection alignment (Fig. 1), because it is easier to build, although transmission alignment can also be used.

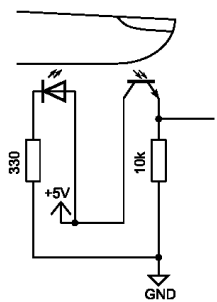

Figure 1. The used reflection arrangement for finger plethysmography.

Due to the reflected light from the tissue, a voltage signal appears on the emitter of the phototransistor, which consists of a larger DC, and a smaller AC component. This AC component is only a few percent of the DC component. An active bandpass filter can be used (1 Hz – 30 Hz) to remove the DC voltage and the high frequency noise [10-11].

Using the microphone input, only alternating voltages in frequency range from 20Hz to 20kHz can be measured. The plethysmograph signals have very low frequency, about 1 – 2 Hz, therefore they can't be connected directly to the microphone input of a sound card. In order to measure such signals, a sort of modulation is needed. Thereupon two methods will be presented.

Since the Raspberry Pi has no microphone input, an external USB sound card should be used. This sort of sound card has low price, hereby it can be procured easily as the Raspberry Pi. The 5 V power supply of both circuits can be taken e. g. from a 5 V USB adapter by USB cable. It is not recommended to supply the measurement circuits from the USB port of the Raspberry Pi, since it is not stable enough.

*A. Amplitude modulation*

Modulation is a technique to measure a DC or low frequency signal by transforming it into a high frequency signal. In case if amplitude modulation, the information is carried by the amplitude of the modulated, high frequency signal. As the block diagram in Fig. 2 shows, our measurement circuit filters, and modulates the signal of the phototransistor. The modulated signal can be connected to the microphone input of the sound card by a 3.5 jack plug. The modulating, *f* frequency signal is provided by the headphone output of the sound card to the measurement circuit via another jack plug.

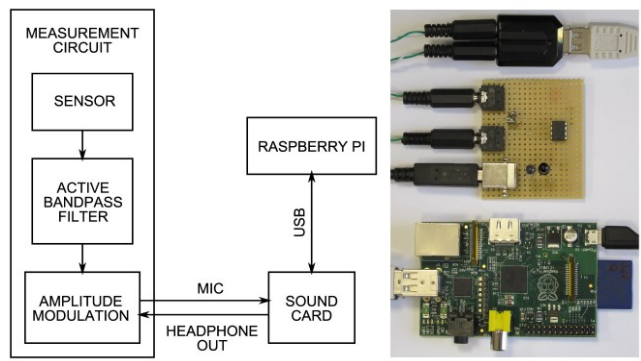

Figure 2. Block diagram of the amplitude modulation measurement technique and the assembled device (above the external sound card, center the measurement circuit, and below the Raspberry Pi).

The amplified and filtered signal is modulated by a sine wave using a transistor, see Fig. 3. The base of this simple amplitude modulation method is that the modulating signal jags the measured signal by the *f* frequency. In plethysmographic measurements, absolute accuracy does not matter, hence the inaccuracy of the sound card causes no error. However, sound cards have good linearity, which is more important.

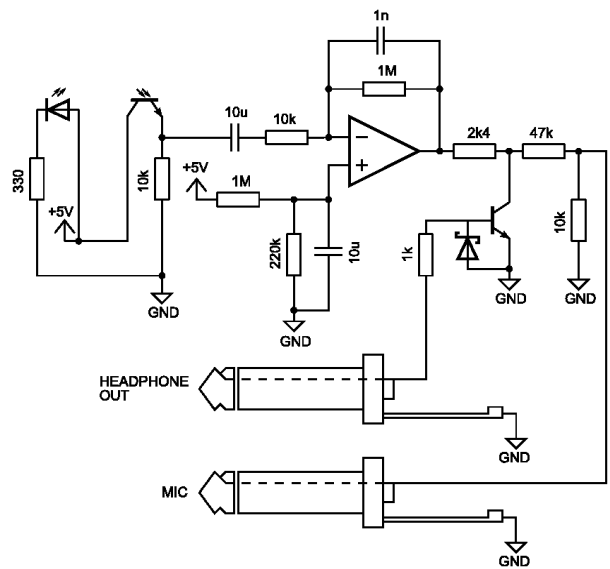

Figure 3. Schematic of the amplitude modulation measurement circuit. The signal of the phototransistor is band-pass filtered, amplified and modulated by a transistor. Finally, it is connected to the microphone input of the sound card, using jack plug. The modulating signal derives from the sound card, also through a jack plug.

The capture of data, demodulation, visualization and further signal processing is done by a software, that we developed in Java and JavaFX. This software also generates the modulating $f$ frequency signal on the headphone output. The used signal processing will be reviewed below.

In demodulation, the Fourier-amplitude of the input signal $x_i$ can be calculated:

$$a_j = \sqrt{\left[\frac{1}{N}\sum_{i=jN}^{(j+1)N-1} x_i \cos\left(2\pi \frac{i}{N}\right)\right]^2 + \left[\frac{1}{N}\sum_{i=jN}^{(j+1)N-1} x_i \sin\left(2\pi \frac{i}{N}\right)\right]^2}$$

, where $N$ is the number of samples in one period in the modulated signal [9]. (We used $f = 1\ kHz$ modulating signal, with $44100\ Hz$ sample rate, then $N = 44$.)

### B. Frequency modulation using an NE555 timer

The bandpass-filtered, low frequency plethysmograph signal also can be transformed by a voltage-to-frequency converter. For that purpose, a timer IC should be used, like the NE 555, which is cheap, common, and easy to use. The voltage, appears on the output of the timer IC, has a higher frequency that is roughly proportional to the amplitude of the plethysmograph signal. That can be connected to the microphone input of the sound card [3,12]. This method is demonstrated by the block diagram in Fig. 4, and the schematic in Fig. 5. We also used the software, which we developed, in the course of frequency modulation measurements. After demodulation, it handles the same way the plethysmograph signals derive from different sources.

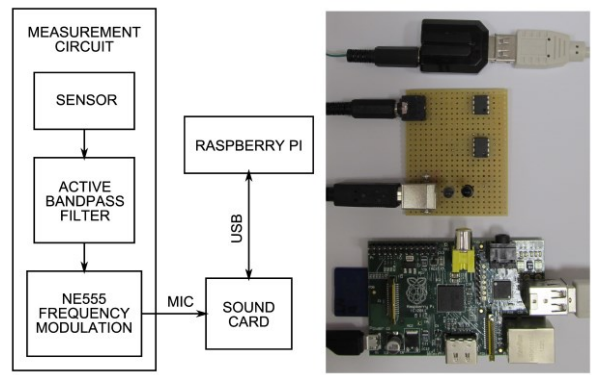

Figure 4. Block diagram of the NE 555 frequency modulation measurement technique and the assembled device (above the external sound card, center the measurement circuit, and below the Raspberry Pi).

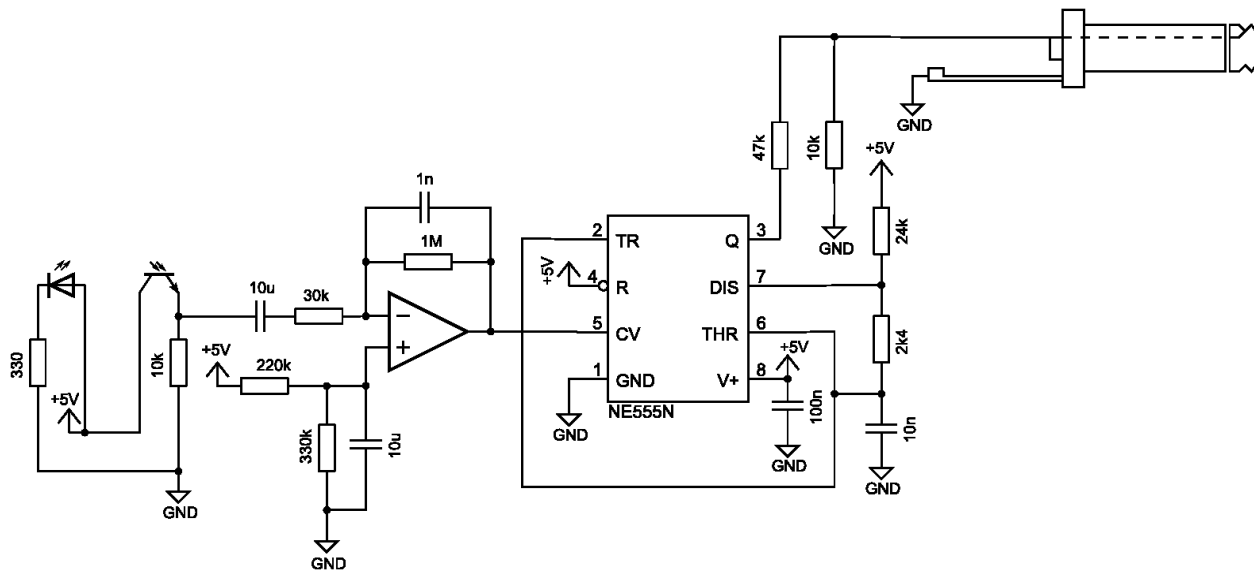

Figure 5. Schematic of the NE 555 frequency modulation measurement circuit. The signal of the phototransistor is band-pass filtered, amplified, and modulated connected to the control voltage pin of the NE 555 timer. The output of the timer IC, which is the modulated signal, is connected to the microphone input of the sound card, using jack plug.

In demodulation, the frequency of the modulated signal should be measured. For that, we used a simple algorithm, which counts the full periods in a fixed time interval [11]. It counts the upward level crossings, $u$ in the measurement window and calculates the time interval appertains to the full periods, $t$, see Fig. 6. Then, the average frequency in the measurement window: $f = (u - 1)/t$. The accuracy of frequency measurement can be improved by using interpolation in the level crossing detection.

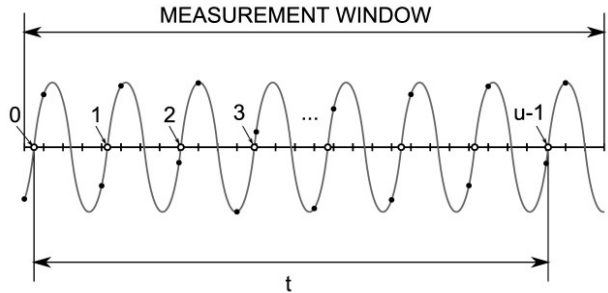

Figure 6. The used method for frequency demodulation. The number of full periods is count in a fix length of time.

III. SIGNAL PROCESSING

The signals, which the demodulation methods result, still can be noisy. The quality of these signals can be enhanced by signal processing. The rate of noise can be decreased in both methods by increasing the length of the measurement window, because then the average is calculated from more data. However, it lowers the time resolution of the signals. Furthermore, to reduce high frequency noise, moving average also can be used.

The signals, conditioned this way are clean enough to detect the heart beats. Before the peaks, owing to heart beats, the rise of the signals has very high maximums. Thus, estimating the derivative by numeric differentiation, these peaks of rise can be easily detected by a level crossing algorithm. This detection could be easier, if the threshold level is adapting to the current signal. Our algorithm uses the maximum value of the numeric derivative curve. The momentary threshold is this maximum multiplied by a constant, which best value was 0.5 for us. After the found rise peak, within a fix time interval, the peak of the plethysmograph signal, or the heart beat presents. It can be detected on this short period of the signal by a maximum searching algorithm.

From the heart beat detection, the time between beats, or the RR intervals and the pulse can be calculated. These values, changes, indicators from additional calculations contain useful information. The three easiest to calculate indicators in a short period are the mean of the pulse, and the mean and the standard deviation of the RR intervals. Additionally, we calculated two other indicators, which are available in shorter time intervals. One of them is the pNN50, the ratio of successive intervals which differ by more than 50 ms and the total number of RR intervals. The other one is the root mean square successive difference (rMSSD) of the RR intervals [5,13].

The software, we developed (see the screenshots on Fig. 7), performs the heart beat detection in real-time, using the method presented above, and shows them on the visualized graph of the plethysmograph signal. Furthermore, it also calculates and visualizes the mentioned indicators.

Although there is no full Java runtime environment with JavaFX, which is capable for the Raspberry Pi yet, an early access is already available [15]. This release is able to run our software on the Raspberry Pi properly. It is not possible to capture screenshots on the Raspberry Pi, because windowing is not included jet in the mentioned package, thus the screenshots in Fig. 7 were made on a PC.

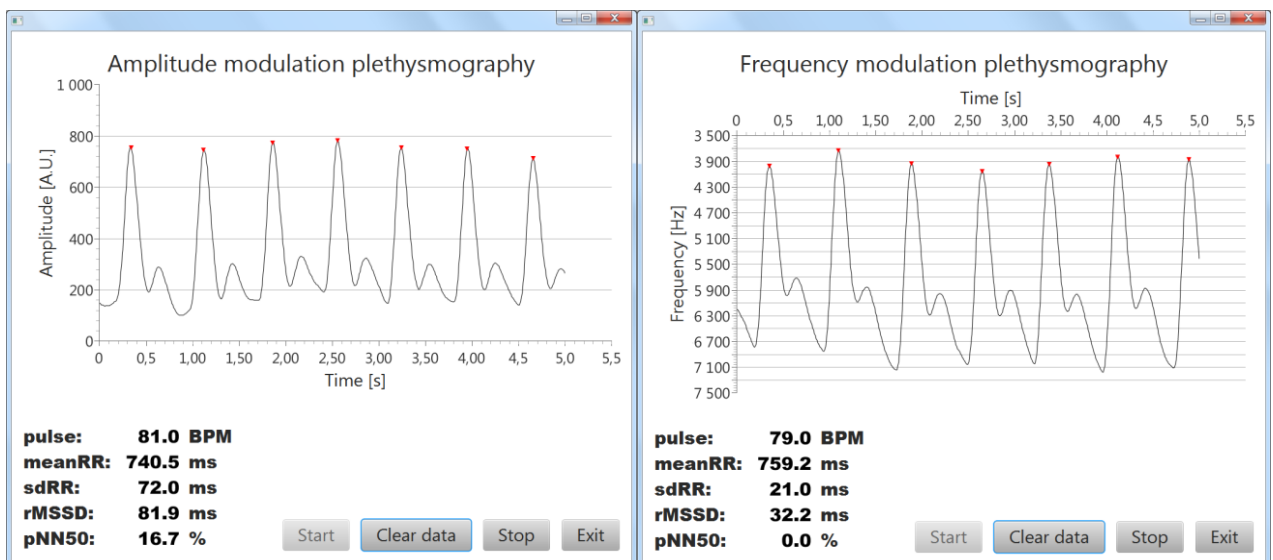

Figure 7. Our software makes the data acquisition, demodulation, moving average and heart beat detection (signed with red triangles) and calculates the mentioned indices. Two screenshots can be seen, with a typical plethysmograph signal, left the result of amplitude modulation, right of frequency modulation. It is yet not possible to make screenshots on the Raspberry Pi of our software, thus these screenshots are from a PC, which runs the same program.

## IV. RESULTS

In this paper, two methods for photoplethysmographic measurements were presented, which are using the Raspberry Pi. It was shown, how to measure these signals by a phototransistor including the whole analog signal conditioning circuitry. Two modulation techniques was presented, which make possible to transmit the plethysmograph signals to an USB sound card. Both amplitude and frequency modulation methods provide high quality plethysmograph signals, but there are some differences in the application of those. One of the advantages of the frequency modulation method against amplitude modulation is that it needs no modulation signal from the sound card, which makes the program and the external circuit simpler. Moreover the frequency of a signal can be measured more exactly on the sound card, then the amplitude. Thus the frequency modulation method is more accurate and reliable. In addition the amplitude inaccuracy of the sound card causes no problems in the amplitude modulation method, because the relative changes are relevant in plethysmography. The precise measurement of these relative changes is insured by the good linearity of the sound card. Furthermore the amplitude modulation method is less noise sensitive, because of the Fourier amplitude demodulation technique. Besides those little differences in precision and ease of use, it is important to note, that the demodulated signal are very similar in many aspects. Additionally we used 10 ms demodulation window in both methods, so the time resolution and the accuracy of peak detection are also the same.

Finally, it was shown, how could signal processing improve the quality of the signals, and which helpful indices could be calculated. The discussed signal processing and further calculations are also done by the software, we developed in Java and JavaFX.

For the presented methods only simple circuits and low-cost means are needed. Notwithstanding the precision of more measured values, e. g. the standard deviation of the RR intervals (sdRR), calculated from peak detection, is close to the precision of the expensive and complicated medical devices. The mentioned sdRR measurement holds a quantitation error caused by the peak detection of a sampled signal. The standard deviation of this quantitation error ($\sigma_q$) is:

$$\sigma_q = \frac{dt}{\sqrt{12}},$$

where $dt$ is the sampling time of the signal. The measured sdRR ($\sigma$) relates to the real sdRR ($\sigma_{RR}$) and the standard deviation of the quantitation error by this equation:

$$\sigma^2 = \sigma_{RR}^2 + \sigma_q^2$$

Thus, if $\frac{\sigma_q^2}{\sigma_{RR}^2} \ll 1$:

$$\frac{\sigma}{\sigma_{RR}} \approx 1 + \frac{1}{2}\frac{\sigma_q^2}{\sigma_{RR}^2}$$

It follows that the relative error in the sdRR, caused by the 100 Hz sampling ($dt = 10$ ms) we used, e. g. for a typical sdRR value of 30 ms, is below 1%.

Besides simplicity and precision, these methods are expandable, and it is possible to improve the used signal processing. Moreover, since the measurements are interfaced by sound card, and the software is written in Java, the presented methods are portable to many other systems.

## V. CONCLUSION

The presented photoplethysmographic measurement methods utilizing the Raspberry Pi are very simple. Furthermore, the used hardware, like the Raspberry Pi, the USB sound card, or the electronic components are low priced and easily procurable (a Raspberry Pi can be purchased for about $25-35, an USB sound cards could be bought from $10, the components of the amplitude modulation, or the frequency modulation circuit cost about $5). In addition, a full documentation of the project is available, and the software developed by us is open source [16]. As a result, these methods can be reproduced easily in a short time, therefore widely available and could be performed even at home. The discussed methods could be used in education, students could learn to use the Raspberry Pi, build such circuits and experiment with the device. In addition, after clinical validation, telemedicine applications could be based on the presented methods. Even though the presented techniques are based on simple and low-cost devices, the obtained precision is corresponding to the execution of RR interval measurements. These methods are also portable to other devices. For personal computers, these work without modification, and could be ported easily e. g. to smart phones, and the BeagleBone Black.

It is important to note, that the external circuits, and signal processing makes these methods for photoplethysmographic measurements more reliable, than e. g. which use the built-in sensors of smart phones [8].

## ACKNOWLEDGMENT

The publication/presentation is supported by the European Union and co-funded by the European Social Fund. Project title: "Telemedicine-focused research activities on the field of Mathematics, Informatics and Medical sciences" Project number: TÁMOP-4.2.2.A-11/1/KONV-2012-0073.